\documentclass[preprint,showpacs,preprintnumbers,amsmath,amssymb,showkeys]{revtex4}
 
\usepackage{graphicx}
\usepackage{dcolumn}
\usepackage{bm}
\usepackage{braket}

\begin{document}

\noindent

\preprint{}

\title{Efficient classical computation of expectation values in a class of quantum circuits with an epistemically restricted phase space representation}
 
\author{Agung Budiyono$^{1,2,3,4}$}
\email{agungbymlati@gmail.com}
\author{Hermawan K. Dipojono$^{1,3}$}
\affiliation{$^1$Research Center for Nanoscience and Nanotechnology, Bandung Institute of Technology, Bandung, 40132, Indonesia}
\affiliation{$^2$Kubus Computing and Research, Juwana, Pati, 59185 Indonesia}
\affiliation{$^3$Department of Engineering Physics, Bandung Institute of Technology, Bandung, 40132, Indonesia}
\affiliation{$^4$Edelstein Center, Hebrew University of Jerusalem, Jerusalem, 91904 Israel} 

\date{\today}
   
\begin{abstract}    
 
We devise a classical algorithm which efficiently computes the quantum expectation values arising in a class of continuous variable quantum circuits wherein the final quantum observable | after the Heisenberg evolution associated with the circuits | is at most second order in momentum. The classical computational algorithm exploits a specific epistemic restriction in classical phase space which directly captures the quantum uncertainty relation, to transform the quantum circuits in the complex Hilbert space into classical albeit unconventional stochastic processes in the phase space. The resulting multidimensional integral is then evaluated using the Monte Carlo sampling method. The work shows that for the specific class of computational schemes, Wigner negativity is not a sufficient resource for quantum speedup. It highlights the potential role of the epistemic restriction as an intuitive conceptual tool which may be used to study the boundary between quantum and classical computations.     
  
\end{abstract}    

\maketitle 

\section{Introduction}

It is widely believed that quantum computers can perform computational tasks exponentially more efficient than classical computers, such as for simulating quantum many body systems \cite{Feynman quantum computer,Lloyd quantum simulator}, or factoring large integer \cite{Shor factorization algorithm}. A series of remarkable studies in the last decades however also revealed that quantum computational algorithms, even those that invoke a lot of peculiar quantum effects such as entanglement, do not always prevail. The first important result in this direction is the Gottesman-Knill theorem \cite{Gottesman-Knill theorem,Nielsen-Chuang book,Aaronson GK theorem improved}, which states that a class of  quantum algorithms which start from a stabilizer state, followed by the operations of Clifford logic gates | which can generate a large amount of entanglement |, can be simulated efficiently on classical computers. This seminal result was further extended to different classes of quantum algorithms \cite{Bartlett continuous variable GK theorem,Bartlett efficient simulation Gaussian quantum optics,Veitch efficient simulation nonnegative WF 1,Veitch efficient simulation nonnegative WF 2,Mari efficient simulation nonnegative WF,Pashayan efficient estimation nonnegative quasiprob,Rahemi-Keshari efficient simulation quantum optics,Valiant efficient simulation Match gates,Knill efficient simulation fermionic linear optics,Terhal efficient simulation fermionic linear optics,Jozsa efficient simulation match gates,Vidal efficient simulation small entanglement,Jozsa efficient simulation small entanglement,Shi efficient simulation small entanglement,Nest MBQC efficient simulation small entanglement,Gross MBQC efficient simulation too much entanglement,Yoran efficient simulation limited width circuits,Markov efficient simulation limited width circuits,Jozsa efficient simulation limited width circuits,Garcia-Alvarez efficient simulation of QC with large negative Wigner function}. Identifying quantum algorithms that can be efficiently simulated classically, and characterizing the distinctive quantum features that are lacking in them, may offer insight to specify the elusive physical resources responsible for the quantum computational speedup \cite{Veitch resource theory quantum computation,Howard contextuality magic,Delfosse contextuality as resource,Vedral quantum computational resource}, and therefore are crucial to better understand the boundary and correspondence between the quantum and classical computations. In turn, deeper grasp of these fundamental problems may provide useful physical intuition to devise new quantum algorithms which outperform their classical counterparts, and to study the minimal requirement for nonuniversal but significantly easier to implement quantum computational schemes, which nevertheless exhibit quantum speedup \cite{Aaronson BosonSampling,Bremmer quantum supremacy IQP,Harrow quantum supremacy review,Lund quantum supremacy review,Boixo characterizing quantum supremacy}.
 
Clearly, the above basic questions in quantum computation are fundamentally linked to the longstanding foundational problem of quantum-classical correspondence and contrast. What are the deep physical concepts which uniquely define quantum mechanics relative to classical mechanics? Or, what is lacking in classical mechanics relative to quantum mechanics? Is it possible to obtain the latter by modifying the former, supplementing it with the necessary conceptual patches? To this end, it has been shown that a significant large set of phenomena traditionally seen as specifically quantum could in fact be explained within classical statistical models with some kinds of ``epistemic (statistical) restrictions'' \cite{Hardy epistricted model,Emerson PhD thesis,Spekkens toy model with epistemic restriction,Bartlett reconstruction of Gaussian QM with epistemic restriction}. Partly inspired by these remarkable results, recently, we have developed a novel phase space representation of quantum mechanics, showing that its opaque formalism in the complex Hilbert space can be phrased more intuitively as a classical statistical mechanics of ensemble of trajectories subjected to a specific epistemic restriction parameterized by a global-nonseparable variable fluctuating randomly on the order of Planck constant \cite{Agung-Daniel model,Agung ERPS representation}. The epistemic restriction captures the quantum uncertainty relation between noncommuting (quantum) positon and momentum operators acting on the abstract Hilbert space in terms of a more intuitive statistical constraint on the allowed distributions over the (classical) phase space. It was further argued that the new phase space formalism can be interpreted as a calculus for estimation of momentum given information on the conjugate positions under the epistemic restriction \cite{Agung epistemic interpretation}, respecting a plausible inferential-causality principle of estimation independence \cite{Agung estimation independence,Agung nonlinear Schroedinger equation and estimation independence} .  

Here, guided by the intuition offered by the epistemically restricted (ER) phase space representation, we devise a classical algorithm which efficiently computes the quantum expectation values arising in a class of continuous variable (CV) quantum computational circuits or CV quantum processes widely encountered in quantum optical settings \cite{Lloyd continuous variable UQC,Gottesman continuous variable UQC}. The computation of quantum expectation value is also a crucial generic step in variational quantum algorithm \cite{Cerezo review VQA}. The idea is to employ the ER phase space representation to transform the quantum circuits into unconventional classical stochastic processes and evaluate the resulting multidimensional integral using the Monte Carlo sampling technique. The classical algorithm thus goes along the spirit of those based on the quasiprobability representations \cite{Hillery quasiprobability review,Lee quasiprobability review,Ferrie quasiprobability review} reported in Refs. \cite{Veitch efficient simulation nonnegative WF 1,Veitch efficient simulation nonnegative WF 2,Mari efficient simulation nonnegative WF,Pashayan efficient estimation nonnegative quasiprob,Rahemi-Keshari efficient simulation quantum optics}. However, unlike the latter, our classical algorithm applies only when the final quantum observable after the Heisenberg evolution associated with the quantum circuits, is at most second order in momentum. The classical simulations based on the quasiprobability representations require that the quasiprobability distributions associated with the quantum circuits are nonnegative  \cite{Veitch efficient simulation nonnegative WF 1,Veitch efficient simulation nonnegative WF 2,Mari efficient simulation nonnegative WF,Rahemi-Keshari efficient simulation quantum optics}; or, that the amount of negative value is sufficiently small to guarantee a fast convergence rate \cite{Pashayan efficient estimation nonnegative quasiprob}. By contrast, our classical algorithm does not suffer from, and thus is not limited by, such problem of negative `probability'. The results therefore show that, for the specific class of computational circuits, Wigner negativity is not a sufficient resource for quantum speedup. A similar suggestion based on different results and method is reported recently in Ref. \cite{Garcia-Alvarez efficient simulation of QC with large negative Wigner function}.     

\section{Epistemically restricted phase space representation of quantum mechanics} 

We summarise the phase space representation of quantum mechanics based on the ER ensemble of trajectories proposed in Refs. \cite{Agung-Daniel model,Agung ERPS representation}. Consider a system of $N$ spatial degrees of freedom $q=(q_1,\dots,q_N)^{\rm T}$, arranged as $N-$elements vector in ${\mathbb R}^N$ where the superscript ${\rm T}$ denotes transposition, with the corresponding canonical conjugate momentum $p=(p_1,\dots,p_N)^{\rm T}$. In classical mechanics, for a system with a classical Hamiltonian $H(q,p)$, the dynamics of the system is deterministic governed by the Hamilton-Jacobi equation, i.e., 
\begin{subequations}\label{Hamilton-Jacobi theory}
\begin{align}
p_n(q;t)&=\partial_{q_n}S_{\rm C},\label{Hamilton-Jacobi condition}\\
-\partial_tS_{\text{C}}(q;t)&=H(q,p), \label{Hamilton-Jacobi equation}
\end{align}
\end{subequations}
$n=1,\dots,N$, where $S_{\rm C}(q;t)$ is a real-valued function of positions $q$ and time $t$, known as the Hamilton's principal function. The Hamilton-Jacobi equation is formally equivalent to the Hamilton's equation \cite{Rund book: Hamilton-Jacobi formalism}. Moreover, the Hamilton-Jacobi equation offers a geometrical picture of an ensemble of trajectories in configuration space: it describes the dynamics of the whole ensemble of trajectories, all characterized by a single Hamilton's principal function $S_{\rm C}(q;t)$. Solving Eq. (\ref{Hamilton-Jacobi theory}) in terms of $S_{\rm C}(q;t)$, a single trajectory in configuration space is picked up by choosing a configuration $q$ at any particular time $t$, and the momentum along the trajectory is obtained by computing Eq. (\ref{Hamilton-Jacobi condition}).  

In classical statistical mechanics of ensemble of trajectories, given $S_{\rm C}(q)$, the probability distribution of $p$ conditioned on $q$ thus reads as, noting Eq. (\ref{Hamilton-Jacobi condition}),  
\begin{eqnarray}
{\mathbb P}_{S_{\rm C}}(p|q)=\prod_{n=1}^{N}\delta\big(p_n-\partial_{q_n}S_{\rm C}(q)\big), 
\label{classical epistemic freedom}
\end{eqnarray} 
(trivial time dependence is implicit). Here and below, a subscript  $F$ in ${\mathbb P}_{F}$ indicates a dependence of the probability on a function $F$. For later convenience, we denote the probability distribution of $q$ at time $t$ with a specific notation as $\rho(q;t)$. The phase space distribution is therefore given by ${\mathbb P}_{\{S_{\rm C},\rho\}}(p,q)={\mathbb P}_{S_{\rm C}}(p|q)\rho(q)$, so that the average of any physical quantity $O(q,p)$ over the phase space distribution can be written as
\begin{eqnarray}
\braket{O}_{\{S_{\rm C},\rho\}}&\doteq&\int{\rm d}^Nq~{\rm d}^Np~O(q,p){\mathbb P}_{S_{\rm C}}(p|q)\rho(q)\nonumber\\
&=&\int{\rm d}^Nq~O(q,\partial_qS_{\rm C})\rho(q), 
\label{classical ensemble average}
\end{eqnarray}
where ${\rm d}^{N}q\doteq{\rm d}q_1\dots{\rm d}q_N$, ${\rm d}^{N}p\doteq{\rm d}p_1\dots{\rm d}p_N$, and we have used Eq. (\ref{classical epistemic freedom}) to get the second line. 

Note that since the classical dynamics governed by Eq. (\ref{Hamilton-Jacobi theory}) conserves the energy along each single trajectory, it automatically conserves the average energy over the whole ensemble of trajectories, i.e., $({\rm d}/{\rm d}t)\langle H\rangle_{\{S_{\rm C},\rho\}}=0$. Conversely, it has been shown in \cite{Agung-Daniel model} that assuming a momentum field which can be written as in Eq. (\ref{Hamilton-Jacobi condition}), and imposing the conservation of average energy and conservation of trajectories (probability current) manifested by the continuity equation $\partial_t\rho+\partial_q\cdot(\rho\dot{q})=0$ where $\dot{q}\doteq{\rm d}q/{\rm d}t$, singles out the Hamilton-Jacobi equation of Eq. (\ref{Hamilton-Jacobi equation}). 

Next, suppose that solving the Hamilton-Jacobi equation, or more straightforwardly solving the equivalent Hamilton's equation, the initial position and momentum are mapped at some later time as: $q\mapsto q'=f_{\rm C}(q,p)$ and $p\mapsto p'=g_{\rm C}(q,p)$, where $f_{\rm C}=(f_{{\rm C}_1},\dots,f_{{\rm C}_N})^{\rm T}$ and $g_{\rm C}=(g_{{\rm C}_1},\dots,g_{{\rm C}_N})^{\rm T}$ are vectors of functions on phase space. This deterministic mapping can be expressed in terms of transition probability as ${\mathbb P}_{\{f_{\rm C},g_{\rm C}\}}(p',q'|p,q)=\prod_{n=1}^N\delta\big(q'_n-f_{{\rm C}_n}(q,p)\big)\prod_{m=1}^N\delta\big(p'_m-g_{{\rm C}_m}(q,p)\big)$. The statistical average of a physical quantity $O(q,p)$ over the phase space distribution after the time evolution can thus be computed as   
\begin{eqnarray}
\braket{O}_{\{(f_{\rm C},g_{\rm C});(S_{\rm C},\rho)\}}&\doteq&\int{\rm d}^Nq'~{\rm d}^Np'~{\rm d}^Nq~{\rm d}^Np~O(q',p'){\mathbb P}_{\{f_{\rm C},g_{\rm C}\}}(p',q'|p,q){\mathbb P}_{\{S_{\rm C},\rho\}}(p,q)\nonumber\\
&=&\int{\rm d}^Nq'~{\rm d}^Np'~{\rm d}^Nq~{\rm d}^Np~O(q',p')\nonumber\\
&\times&\prod_{n=1}^N\delta\big(q'_n-f_{{\rm C}_n}(q,p)\big)\prod_{m=1}^N\delta\big(p'_m-g_{{\rm C}_m}(q,p)\big)\prod_{k=1}^N\delta(p_k-\partial_{q_k}S_{\rm C}(q))\rho(q).
\label{classical ensemble average time evolution}
\end{eqnarray} 

We have argued in Refs. \cite{Agung-Daniel model,Agung ERPS representation} that the abstract formulas of quantum mechanics can be expressed as a specific modification of the above classical statistical mechanics of ensemble of trajectories in phase space. First, we introduce an ``ontic extension'' in the form of a global-nonseparable variable $\xi$: it is real-valued with the dimension of action and depends only on time (i.e., spatially uniform). We assume that $\xi$ fluctuates randomly on a microscopic time scale with a probability density $\chi(\xi)$ such that the first and second moments are constant in time, given by 
\begin{equation}
\overline{\xi}\doteq\int {\rm d}\xi~\xi~\chi(\xi)=0,~~~\overline{\xi^2}=\hbar^2.
\label{Planck constant}
\end{equation} 
 
We then impose a specific epistemic restriction on the class of phase-space probability distributions that Nature allows us to prepare, as follows. Consider an ensemble of identical preparations (characterized by the same set of controllable macroscopic parameters) resulting in an ensemble of trajectories in configuration space following a momentum field. In conventional classical statistical mechanics, from  Eq. (\ref{Hamilton-Jacobi condition}), it is possible to prepare an ensemble of trajectories with a given targeted probability distribution of positions $\rho(q)$ using an arbitrary form of momentum field $p(q)$ by choosing an arbitrary $S_{\rm C}(q)$. This means that, in classical mechanics, as shown in Eq. (\ref{classical epistemic freedom}), the probability distribution of $p$ conditioned on $q$ is independent of $\rho(q)$, i.e., ${\mathbb P}_{\{S_{\rm C},\rho\}}(p|q)={\mathbb P}_{S_{\rm C}}(p|q)$. Or, equivalently, a given momentum field can be used to prepare an ensemble of trajectories with an arbitrary targeted $\rho(q)$, i.e., each trajectory in the momentum field can be weighted with an arbitrary choice of $\rho(q)$. 

Let us assume that such an ``epistemic (statistical) freedom'' in conventional classical statistical mechanics is no longer fully granted in microscopic world. Hence, we assume that in microscopic regime, each trajectory in a given momentum field can no longer be assigned an arbitrary weight $\rho(q)$, i.e., the allowed distribution of positions is fundamentally restricted by the underlying momentum field \cite{Agung epistemic interpretation,Agung estimation independence}. Or, equivalently, given a targeted $\rho(q)$, it is no longer possible to prepare an ensemble of trajectories realizing $\rho(q)$ following an arbitrary momentum field. This means that the probability distribution of $p$ conditioned on $q$ must depend on the targeted $\rho(q)$, i.e.: ${\mathbb P}_{\{\dots,\rho\}}(p|q,\dots)\neq{\mathbb P}_{\{\dots\}}(p|q,\dots)$. Let us then consider a statistical model with such an epistemic restriction so that an ensemble of identical preparations yields a probability distribution of $p$ conditioned on $q$ which fundamentally depends on $\rho(q)$ as follows (Cf. Eq. (\ref{classical epistemic freedom})) \cite{Agung-Daniel model}: 
\begin{equation}  
{\mathbb P}_{\{S ,\rho\}}(p|q,\xi)=\prod_{n=1}^N\delta\Big(p_n-\Big(\partial_{q_n}S +\frac{\xi}{2}\frac{\partial_{q_n}\rho}{\rho}\Big)\Big),   
\label{fundamental epistemic restriction} 
\end{equation}
parameterized by the global random variable $\xi$ satisfying Eq. (\ref{Planck constant}). Here, $S (q;t)$ is a real-valued scalar function of $(q;t)$ with the dimension of action, replacing the role of $S_{\rm C}(q;t)$ in Eq. (\ref{classical epistemic freedom}). From Eq. (\ref{fundamental epistemic restriction}), the allowed (conditional) phase space distributions are thus also restricted to have the following specific form:      
\begin{eqnarray}
{\mathbb P}_{\{S,\rho\}}(p,q|\xi)&=&{\mathbb P}_{\{S,\rho\}}(p|q,\xi)\rho(q)\nonumber\\
&=&\prod_{n=1}^N\delta\Big(p_n-\Big(\partial_{q_n}S +\frac{\xi}{2}\frac{\partial_{q_n}\rho}{\rho}\Big)\Big)\rho(q). 
\label{ER phase space distribution}
\end{eqnarray}

One can straightforwardly show that Eqs. (\ref{fundamental epistemic restriction}) and (\ref{Planck constant}) imply the Heisenberg-Kennard uncertianty relation, i.e., $\sigma_{q_n}\sigma_{p_n}\ge\hbar/2$, $n=1,\dots,N$, where $\sigma_{q_n}$ and $\sigma_{p_n}$ are respectively the standard deviations of position and momentum of the $n$th spatial degree of freedom \cite{Agung-Daniel model}. In this sense, the epistemic restriction constrains the allowed phase space distributions to satisfy the uncertainty relation. For this reason, we refer to ${\mathbb P}_{\{S,\rho\}}(p,q|\xi)$ defined in Eq. (\ref{ER phase space distribution}) as the ``ER (epistemically restricted) phase space distribution'' associated with a pair of real-valued functions $(S,\rho)$. We have recently shown in Refs. \cite{Agung ERPS representation,Agung epistemic interpretation} that the ER phase space distribution is not just a mathematical artefact, but can be indirectly reconstructed in experiment via the notion of weak momentum value defined as $\frac{\braket{q|\hat{p}|\psi}}{\braket{q|\psi}}$ \cite{Aharonov weak value,Lundeen complex weak value,Jozsa complex weak value}. This, in turn, provides an interpretation of the complex weak momentum value in terms of the ER momentum fluctuation and quantum uncertainty. The classical limit of macroscopic physical regime is obtained when $|\partial_qS |\gg|\frac{\xi}{2}\frac{\partial_q\rho}{\rho}|$, e.g., $|\xi|\rightarrow 0$ or equivalently $\hbar\rightarrow 0$ as per Eq. (\ref{Planck constant}), and $S \rightarrow S_{\rm C}$, so that the conditional distribution of momentum of Eq. (\ref{fundamental epistemic restriction}) reduces back to that in conventional classical statistical mechanics given by Eq. (\ref{classical epistemic freedom}), and the ER phase space distribution of Eq. (\ref{ER phase space distribution}) rolls back to the classical phase space distribution as ${\mathbb P}_{\{S,\rho\}}(p,q|\xi)\rightarrow {\mathbb P}_{\{S_{\rm C},\rho\}}(p,q)=\prod_{n=1}^N\delta(p_n-\partial_{q_n}S_{\rm C})\rho(q)$, lacking an epistemic restriction. 

Assume further that the physical quantities in the above ER classical model are given by some real-valued functions of positions and momentum, $O(q,p)$, having the same form as those in classical mechanics. We then obtain the following theorem expressing the equivalence between the conventional statistical average over the ER phase space distribution and the quantum expectation value.\\
{\bf Theorem 1} (Budiyono-Rohrlich \cite{Agung-Daniel model}):\\
Consider an ensemble of trajectories satisfying the epistemic restriction of Eqs. (\ref{fundamental epistemic restriction}) and (\ref{Planck constant}). The ensemble average of any classical physical quantity $O(q,p)$ up to second order in momentum $p$ over the ER phase space distribution of Eq. (\ref{ER phase space distribution}) is equal to the quantum expectation value as 
\begin{eqnarray}
\braket{O}_{\{S ,\rho\}}&\doteq&\int {\rm d}^Nq~{\rm d}\xi~{\rm d}^Np~O(q,p){\mathbb P}_{\{S ,\rho\}}(p,q|\xi)\chi(\xi)\nonumber\\
&=&\int {\rm d}^Nq~{\rm d}\xi~{\rm d}^Np~O(q,p)\prod_{n=1}^N\delta\Big(p_n-\Big(\partial_{q_n}S +\frac{\xi}{2}\frac{\partial_{q_n}\rho}{\rho}\Big)\Big)\rho(q)\chi(\xi)\nonumber\\
&=&\braket{\psi|\hat{O}|\psi}, 
\label{equivalence principle}
\end{eqnarray}
where $\hat{O}$ is a Hermitian operator taking exactly the same form as that obtained by applying the Dirac canonical quantization procedure to $O(q,p)$ with a specific ordering of operators, and the wave function $\psi(q;t)\doteq\langle q|\psi(t)\rangle$ is defined as    
\begin{equation} 
\psi(q;t)\doteq\sqrt{\rho(q;t)}\exp(iS(q;t)/\hbar). 
\label{wave function}
\end{equation} 
See the Methods section in Ref. \cite{Agung-Daniel model} for the proof. We note that when the physical quantity $O(q,p)$ has cross terms between momentum of different degrees of freedom,  i.e., $p_np_m$, $n\neq m$, the nonseparibility of $\xi$ is indeed indispensable for obtaining Eq. (\ref{equivalence principle}). 

A couple of important notes are in order here. Theorem 1 is developed in such a way that we can devise a classical statistical model so that the quantum expectation value is `reconstructed' from the conventional statistical average over the ER phase space distribution of Eq. (\ref{ER phase space distribution}). We can however read Theorem 1 the other way around. Namely, the ER classical model can be seen as a `representation' of quantum statistics in the ER classical phase space \cite{Agung ERPS representation}. In this reading, first we are given a quantum wave function $\psi(q)$ describing the preparation. We compute its amplitude $\rho(q)$ and phase $S(q)$ as
\begin{eqnarray}
\rho(q)=\big|\psi(q)\big|^2~~~ \&~~~S (q)=\frac{\hbar}{2i}\big(\log\psi(q)-\log\psi(q)^*\big),
\label{phase and amplitude}
\end{eqnarray}
complying with Eq. (\ref{wave function}), based on which, we define the ER phase space distribution of Eq. (\ref{ER phase space distribution}). We then use the ER phase space distribution to transform the quantum expectation value into a conventional statistical phase space average as in Eq. (\ref{equivalence principle}) | by reading it from the right-hand side to the left. 

Reading Eq. (\ref{equivalence principle}) as an epistemically restricted phase space representation of quantum expectation values, let us briefly discuss the apparent asymmetric treatment of $q$ and $p$ in the Theorem 1. Namely, while there is no restriction on the order of $q$, Theorem 1 only applies when $p$ is at most second order. This asymmetry seems in particular peculiar from the perspective of quantum optics, wherein the quadrature phase space operators $(\hat{q},\hat{p})$ denote the field amplitudes oscillating out of phase with each other, so that they can switch positions. First, we emphasize that, as in quantum optics, the labelling of $q$ and $p$ in Theorem 1 is a matter of convention; namely, we can interchange the roles of the symbol $q$ and $p$ by working in the $p$ (i.e., momentum) representation instead of in the $q$ (i.e., position) representation. For a concrete example, consider the computation of the following expectation value $\braket{\psi|(\hat{q}^2+\hat{p}^n)|\psi}$, where $n$ is an integer larger than two. In this case, it is natural to work in the momentum representation and write the associated wave function in the polar form as $\phi(p)\doteq\braket{p|\psi}=\sqrt{\rho(p)}e^{iS(p)/\hbar}$. One can then show that Eq. (\ref{equivalence principle}) in Theorem 1 still applies with the transformation $q\leftrightarrow p$. However, in this case, $\hat{q}$ must be at most second order. Hence, either way, within the ER phase space representation, one of the two quadrature phase space operators must be at most second order. Mathematically, the asymmetric treatment of $q$ and $p$ in Theorem 1 can be traced to their asymmetric roles in the epistemic restriction based on which the phase space representation is constructed. Below we shall take the convention that $p$ plays the role of momentum which is limited to be at most second order. 

As an important example of the application of Theorem 1, we have:
\begin{eqnarray}
\braket{(\eta_m-\braket{\eta_m}_{\{S,\rho\}})(\eta_n-\braket{\eta_n}_{\{S,\rho\}})}_{\{S ,\rho\}}=\braket{\eta_m\eta_n}_{\{S ,\rho\}}-\braket{\eta_m}_{\{S ,\rho\}}\braket{\eta_n}_{\{S ,\rho\}}\nonumber\\
=\braket{\psi|\frac{1}{2}(\hat{\eta}_m\hat{\eta}_n+\hat{\eta}_n\hat{\eta}_m)|\psi}-\braket{\psi|\hat{\eta}_m|\psi}\braket{\psi|\hat{\eta}_n|\psi}, 
\label{covariance of position and momentum variables}
\end{eqnarray}
where $\eta=(p,q)$, $m,n=1,\dots,N$. Notice that the right-hand side is precisely the quantum covariance matrix of the phase space (quadrature) operators. Hence, reading Eq. (\ref{covariance of position and momentum variables}) from the right-hand side to the left, the covariance matrix of position and momentum operators in quantum mechanics, which plays prominent roles in quantum optics and also in general CV quantum information processing, can be expressed as the ordinary classical statistical covariance matrix of the position and momentum over the ER phase space distribution.  

It can be further shown that imposing the conservation of average energy to the ensemble of trajectories satisfying the epistemic restriction of Eqs. (\ref{fundamental epistemic restriction}) and (\ref{Planck constant}), i.e., $({\rm d}/{\rm d}t)\braket{H}_{\{S,\rho\}}=0$, and also the conservation of trajectories (probability current), single out a unique dynamics for the wave function defined in Eq. (\ref{wave function}) given by the unitary Schr\"odinger equation: $i\hbar({\rm d}/{\rm d}t)\ket{\psi}=\hat{H}\ket{\psi}$ \cite{Agung-Daniel model}. $\hat{H}$ is the quantum Hamiltonian having the same form as that obtained by applying the Dirac canonical quantization procedure to $H(q,p)$ with a specific ordering of operators. In the macroscopic regime of classical limit so that Eq. (\ref{fundamental epistemic restriction}) becomes Eq. (\ref{classical epistemic freedom}), Eq. (\ref{equivalence principle}) gives back the conventional classical average of Eq. (\ref{classical ensemble average}), and the Schr\"odinger equation reduces to the classical Hamilton-Jacobi equation of Eq. (\ref{Hamilton-Jacobi theory}) \cite{Agung-Daniel model,Agung epistemic interpretation}.  

\section{Computing the expectation values in a class of quantum circuits with an epistemically restricted ensemble of trajectories}

We extend Theorem 1 to include a specific class of mappings of position and momentum variables to devise classical algorithms which efficiently compute the quantum expectation values arising in a nontrivial class of CV quantum computational circuits or CV quantum processes on classical probabilistic computers. We note that a quantum circuit computing the quantum expectation value constitutes an important generic step in the recently proposed scheme of variational quantum algorithm to address the limitation of the currently available quantum computer with a small number of qubits \cite{Cerezo review VQA}.   

Consider the following stochastic dynamics of ensemble of trajectories. Suppose that at an initial time we are given a pair of real-valued functions $(S(q),\rho(q))$, where $\rho(q)$ is a normalized probability density of $q$. First, we draw a sample of the pair of variables $(q,\xi)$ from the joint probability density $\rho(q)\chi(\xi)$, where $\chi(\xi)$ satisfies Eq. (\ref{Planck constant}). We then use $(q,\xi)$ to compute the value of $p$ at this initial time as 
\begin{eqnarray}
p_n=\partial_{q_n}S(q)+\frac{\xi}{2}\frac{\partial_{q_n}\rho(q)}{\rho(q)}, 
\label{computation of initial momentum}
\end{eqnarray}
$n=1,\dots,N$. Hence, in this way, we initially sample the phase space variables $(p,q)$ from the joint probability distribution of Eq. (\ref{ER phase space distribution}), which is just the ER phase space distribution associated with a wave function $\psi(q)=\sqrt{\rho(q)}\exp(iS(q)/\hbar)$, summarized in the previous section. 

Next, assume that the phase space variables $(p,q)$ evolves in time into a new phase space variable $(p',q')$ as  
\begin{eqnarray}
q\mapsto q'&=&f(q,p),\nonumber\\
p\mapsto p'&=&g(q,p), 
\label{classical deterministic phase-space mapping}
\end{eqnarray}
where $f=(f_1,\dots,f_N)^{\rm T}$ and $g=(g_1,\dots,g_N)^{\rm T}$ are vectors of functions on phase space, independent of $\xi$. Here, we assume for simplicity that $\xi$ is kept fixed during the mapping of phase space variables (i.e., during the time evolution). The above deterministic mapping induces the following transition probability over the phase space variables: 
\begin{eqnarray}
{\mathbb P}_{\{f,g\}}(p',q'|p,q)&=&\prod_{n=1}^N\delta\big(q'_n-f_n(q,p)\big)\prod_{m=1}^N\delta\big(p'_m-g_m(q,p)\big). 
\label{phase space transition probability}
\end{eqnarray}

At the end of the phase space transformation, we wish to compute the statistical phase space average of a physical quantity $O(q',p')$. We therefore have to evaluate the following multidimensional integral:
\begin{eqnarray}
\braket{O}_{\{(f,g);(S ,\rho)\}}&\doteq&\int{\rm d}^Nq'~{\rm d}^Np'~{\rm d}^Nq~{\rm d}\xi~{\rm d}^Np~O(q',p'){\mathbb P}_{\{f,g\}}(p',q'|p,q){\mathbb P}_{\{S,\rho\}}(p,q|\xi)\chi(\xi)\nonumber\\
&=&\int{\rm d}^Nq'~{\rm d}^Np'~{\rm d}^Nq~{\rm d}\xi~{\rm d}^Np~O(q',p')\prod_{n=1}^N\delta\big(q'_n-f_n(q,p)\big)\prod_{m=1}^N\delta\big(p'_m-g_m(q,p)\big)\nonumber\\
&\times&\prod_{k=1}^N\delta\Big(p_k-\Big(\partial_{q_k}S +\frac{\xi}{2}\frac{\partial_{q_k}\rho}{\rho}\Big)\Big)\rho(q)\chi(\xi),
\label{classical computation for ensemble average general}
\end{eqnarray} 
where we have used Eqs. (\ref{ER phase space distribution}) and (\ref{phase space transition probability}) to get the second equality. Note that the above computational scheme for average value can be seen as a classical stochastic process for $2N$ positions and momentum random variables, but with an initial phase space that is epistemically (statistically) restricted being sampled from the specific ER phase space distribution given by Eq. (\ref{ER phase space distribution}). Hence, assuming that the time evolution of the position and momentum variables described in Eq. (\ref{classical deterministic phase-space mapping}) is efficiently tractable on a classical computer, and provided we can sample $\xi$ from $\chi(\xi)$, and $(p,q)$ from ${\mathbb P}_{\{S,\rho\}}(p,q|\xi)$ of Eq. (\ref{ER phase space distribution}) efficiently, the above computational task can be run on a classical computer efficiently using Monte Carlo sampling method \cite{Nest probabilistic method}. We show below that for a specific class of classical quantities $O(q,p)$, and phase space mappings $(f,g)$, Eq. (\ref{classical computation for ensemble average general}) can be used to efficiently compute the expectation values arising in a wide important class of quantum circuits. 

First, consider the case when $f$ and $g$ in Eq. (\ref{classical computation for ensemble average general}) are linear in position and momentum variables; namely, we assume the following linear mapping of position and momentum variables:
\begin{eqnarray}
q'&=&f(q,p)=A\cdot q+B\cdot p+ I\cdot q_0,\nonumber\\
p'&=&g(q,p)=C\cdot q+D\cdot p+ I\cdot p_0,
\label{linear mappings of phase space variables} 
\end{eqnarray}
where $A$, $B$, $C$ and $D$ are $N\times N$-matrices, $I$ is identity matrix, $q_0$ and $p_0$ are $N$-elements column vectors, and $(A\cdot q)_m\doteq \sum_{n=1}^NA_{mn}q_n$, et cetera. Inserting Eq. (\ref{linear mappings of phase space variables}) into Eq. (\ref{classical computation for ensemble average general}), and evaluating the integration over $(p',q')$, we obtain 
\begin{eqnarray}
\braket{O}_{\{(f,g);(S ,\rho)\}}&=&\int{\rm d}^Nq~{\rm d}\xi~{\rm d}^Np~O'(q,p)\prod_{k=1}^N\delta\Big(p_k-\Big(\partial_{q_k}S +\frac{\xi}{2}\frac{\partial_{q_k}\rho}{\rho}\Big)\Big)\rho(q)\chi(\xi). 
\label{linear optics1}
\end{eqnarray}  
where $O'(q,p)\doteq O\big(A\cdot q+B\cdot p+ I\cdot q_0,C\cdot q+D\cdot p+ I\cdot p_0\big)$. 

Now, let us assume that $O(q',p')$ in Eq. (\ref{classical computation for ensemble average general}) is at most quadratic in the phase space variables $(p',q')$, i.e., it may contain terms like $p'_mp'_n$, $q'_mq'_n$, or $p'_mq'_n$, but it does not contain cubic or higher order terms like $q_mq_nq_l$, $m,n,l=1,\dots,N$. In this case, imposing the linear transformation of Eq. (\ref{linear mappings of phase space variables}), the resulting transformed $O'(q,p)$ in Eq. (\ref{linear optics1}) must also be at most second order in $(p,q)$. Importantly, $O'(q,p)$ is therefore at most second order in $p$. This fact permits the application of Theorem 1 to show that the statistical average over the ER phase space distribution of Eq. (\ref{linear optics1}) is equivalent to the quantum expectation value as  
\begin{eqnarray}
\braket{O}_{\{(f,g);(S ,\rho)\}}=\braket{\psi|O'(\hat{q},\hat{p})|\psi}. 
\label{linear optics 2}
\end{eqnarray}
Here, the wave function $\psi$ is defined as in Eq. (\ref{wave function}), and the quantum observable is given by $O'(\hat{q},\hat{p},)\doteq O\big(A\cdot\hat{q}+B\cdot\hat{p}+\hat{I}\cdot q_0,C\cdot\hat{q}+D\cdot\hat{p}+\hat{I}\cdot p_0\big)$. Hence, $O'(\hat{q},\hat{p},)$ is obtained from $O(\hat{q},\hat{p})$ via the following linear Affine transformation of the position and momentum operators: 
\begin{eqnarray}
\hat{q}'&=&f(\hat{q},\hat{p})=A\cdot\hat{q}+B\cdot\hat{p}+\hat{I}\cdot q_0,\nonumber\\
\hat{p}'&=&g(\hat{q},\hat{p})=C\cdot\hat{q}+D\cdot\hat{p}+\hat{I}\cdot p_0.
\label{linear Affine transformation}
\end{eqnarray}

Suppose further that the linear transformation of the position and momentum variables of Eq. (\ref{linear mappings of phase space variables}) is symplectic \cite{Arvind symplectic geometry}; namely, it preserves the canonical Poisson bracket relations: $[q'_n,p'_m]_{\rm PB}=[q_n,p_m]_{\rm PB}=\delta_{nm}$, $n,m=1,\dots,N$. Then, the corresponding Affine transformation of the position and momentum operators of Eq. (\ref{linear Affine transformation})  conserves the canonical commutation relations, i.e., $[\hat{q}_n',\hat{p}_m']=[\hat{q}_n,\hat{p}_m]=i\hbar\delta_{nm}$, $n,m=1,\dots,N$, so that the mapping $O(\hat{q},\hat{p})\mapsto O'(\hat{q},\hat{p})$ can be implemented by a unitary transformation $\hat{U}_{\{f,g\}}$ as \cite{Arvind symplectic geometry,Adesso CV quantum information}
\begin{eqnarray}
O'(\hat{q},\hat{p})=\hat{U}^{\dagger}_{\{f,g\}}O(\hat{q},\hat{p})\hat{U}_{\{f,g\}}.
\label{Heisenberg equation}
\end{eqnarray}  
Here $\hat{U}_{\{f,g\}}\doteq\exp(-i\hat{H}\theta/\hbar)$, where $\hat{H}$ | the quantum Hamiltonian | is the generator of the unitary transformation taking the form of a Hermitian operator at most quadratic in position and momentum operators, and $\theta$ is a parameter with the dimension of time. Such unitaries are called Gaussian unitaries since they map Gaussian states onto Gaussian states: they only change the means and the covariances of the initial Gaussian states. The set of all Gaussian unitaries $\hat{U}_{\{f,g\}}$ comprises a Clifford group and plays crucial roles in quantum optics and in general CV quantum information processing \cite{Bartlett continuous variable GK theorem,Bartlett efficient simulation Gaussian quantum optics,Adesso CV quantum information,Weedbrook Gaussian quantum information}.    

Noting Eq. (\ref{Heisenberg equation}), Eq. (\ref{linear optics 2}) can thus be written as 
\begin{eqnarray}
\braket{O}_{\{(f,g);(S ,\rho)\}}=\braket{\psi|\hat{U}_{\{f,g\}}^{\dagger}O(\hat{q},\hat{p})\hat{U}_{\{f,g\}}|\psi}. 
\label{simulation of linear quantum optics}
\end{eqnarray}
Reading Eq. (\ref{simulation of linear quantum optics}) from the right-hand side to the left, we therefore have the following theorem.\\
{\bf Theorem 2}:\\
The expectation values of quantum observables up to second order in position and momentum operators arising in any quantum circuits which start from arbitrary quantum states and then acted upon by unitary quantum gates generated by quadratic quantum Hamiltonians (Gaussian unitaries), can be computed efficiently on classical probabilistic computers using Monte Carlo sampling over the ER ensemble of trajectories.  

We have shown that the classical algorithm above can compute the expectation value of quantum observables at most quadratic in position and momentum operators. It thus does not in general simulate the full quantum state. However, for the specific yet important class of quantum circuits which initiate from an arbitrary Gaussian state and evolve under an arbitrary Gaussian unitary inducing linear Affine transformation (in the Heisenberg picture) of the type of Eq. (\ref{linear Affine transformation}), our classical algorithm can be used to simulate the final quantum state. To see this, first recall that the initial Gaussian quantum states are transformed by the Gaussian unitaries into Gaussian states. Furthermore, noting that Gaussian states are completely determined by the means and covariances of the position and momentum operators \cite{Adesso CV quantum information,Weedbrook Gaussian quantum information}, to know the final Gaussian states, we only need to compute their means and covariances, i.e., $\braket{\psi_{\rm G}|\hat{U}_{\rm G}^{\dagger}\hat{\eta}_k\hat{U}_{\rm G}|\psi_{\rm G}}$ and $\braket{\psi_{\rm G}|\hat{U}_{\rm G}^{\dagger}\frac{1}{2}(\hat{\eta}_k\hat{\eta}_l+\hat{\eta}_l\hat{\eta}_k)\hat{U}_{\rm G}|\psi_{\rm G}}$, $\eta=p,q$, and $k,l=1,\dots,N$, where $\ket{\psi_{\rm G}}$ is the initial Gaussian state and $\hat{U}_{\rm G}$ is the Gaussian unitary. The above quantum circuits for computing the quantum averages and quantum covariances of the phase space quadrature operators obviously fall into the scope of Theorem 2 so that they can be efficiently computed using our classical sampling algorithm. Moreover, the means and covariances of the phase space operators over the final Gaussian states can be further used to compute the probability of measurement outcome with respect to the Gaussian POVM. 

We have thus the following corollary of Theorem 2.\\
{\bf Corollary 3}:\\
Quantum algorithms that start from the preparations of Gaussian states, followed by the operations of Gaussian unitaries and measurements over Gaussian POVM, can be simulated efficiently on classical probabilistic computers using Monte Carlo sampling over the ER ensemble of trajectories.\\
Notice that the classical simulatability of Gaussian quantum computations in Corollary 3 is just the CV version of the Gottesman-Knill theorem reported in \cite{Bartlett continuous variable GK theorem,Bartlett efficient simulation Gaussian quantum optics}. Moreover, note that the CV Gottesman-Knill theorem does not apply for the computation of the expectation value when the initial quantum state is not Gaussian. In this sense, our classical algorithm summarized in Theorem 2 therefore extends the CV Gottesman-Knill theorem to include the computations of the expectation values of observables quadratic in position and momentum operators with nonGaussian initial quantum states. 

Theorem 2 can also be extended to cover a class of nonlinear transformations of phase space operators induced by nonGaussian quantum operations as follows. We first note that to prove Theorem 2, we have used Theorem 1 which only requires that $O(\hat{q},\hat{p})$ in Eq. (\ref{equivalence principle}) is at most second order in $\hat{p}$; apart from that, it may contain terms with arbitrary degrees of $\hat{q}$, e.g., $\hat{q}^4$, $\hat{p}\hat{q}^3\hat{p}$, et cetera. This means that we can devise an efficient classical algorithm which computes the quantum expectation values arising in quantum circuits starting from an arbitrary initial quantum state followed by the application of any quantum gate as long as the final quantum observable | after the Heisenberg evolution induced by the quantum circuits | does not have term third order or higher in momentum operators. For example, consider the following quantum circuit for computing the quantum expectation value $\braket{\psi|\hat{U}^{\dagger}_{\{f,g\}}O(\hat{q},\hat{p})\hat{U}_{\{f,g\}}|\psi}$, where $O(\hat{q},\hat{p})=\hat{p}^2+\hat{q}^4$ and $\hat{U}_{\{f,g\}}$ is a nonGaussian quantum gate inducing nonlinear transformations $\hat{p}\mapsto \hat{p}+\hat{q}^2$ and $\hat{q}\mapsto\hat{q}$. One can see that under such  transformation, the quantum observable after the Heisenberg evolution is maintained to be second order in $\hat{p}$: $O(\hat{q},\hat{p})\mapsto O'(\hat{q},\hat{p})=\hat{p}^2+\hat{p}\hat{q}^2+\hat{q}^2\hat{p}+2\hat{q}^4$. Hence, we can still proceed as before to efficiently compute the above quantum expectation value using the classical algorithm of Eq. (\ref{classical computation for ensemble average general}) and evaluate the integral using Monte Carlo sampling over the associated ER phase space distribution.  We therefore obtain the following theorem which is more general than Theorem 2.\\
{\bf Theorem 4}:\\
The quantum expectation values arising in any quantum circuits which start from arbitrary quantum states followed by the applications of unitary quantum gates inducing in general nonlinear transformations (in the Heisenberg picture) and yielding quantum observables at most quadratic in momentum operators, can be classically computed efficiently using Monte Carlo sampling over the ER ensemble of trajectories.    

Let us summarize, before proceeding, the important ingredients for the classical algorithm based on the ER phase space representation of Eq. (\ref{classical computation for ensemble average general}). First, given an initial pure quantum state $\ket{\psi}$, we express it in the configuration representation to get the wave function $\psi(q)=\braket{q|\psi}$, and compute its phase $S(q)$ and amplitude $\rho(q)$ as in Eq. (\ref{phase and amplitude}). Hence, we assume that the wave function $\psi(q)$, its phase $S(q)$, and amplitude $\rho(q)$, can be computed efficiently, and $\rho(q)$ can be sampled efficiently. These requirements give a restriction on the form of the initial quantum states. For discrete variables case, such quantum states are called computationally trackable states in Ref. \cite{Nest probabilistic method}. Moreover, we also assume that the computation of $p$ in Eq. (\ref{computation of initial momentum}), and also the phase space transformation in Eq. (\ref{classical deterministic phase-space mapping}), can be carried out efficiently on classical computers. 

In the macroscopic classical limit, noting that the mapping of Eq. (\ref{classical deterministic phase-space mapping}) is given by the classical Hamilton's equation, and Eq. (\ref{ER phase space distribution}) reduces back to the phase space distribution of conventional classical statistical mechanics, then Eq. (\ref{classical computation for ensemble average general}) gives back the computation of the average value in conventional classical statistical mechanics of Eq. (\ref{classical ensemble average time evolution}). The epistemic restriction of Eq. (\ref{fundamental epistemic restriction}) is thus playing a crucial role in the quantum circuits for computing expectation values. Namely, first, the epistemic restriction is a necessary ingredient for any form of quantum algorithm to compute the quantum expectation value without which we regain the genuine conventional classical algorithm. In other words, any such quantum algorithm (covertly) operates a specific epistemic restriction encoded in the Hilbert space in the form of noncommutative structure of the phase space operators. Moreover, for a specific class of quantum circuits computing the expectation values, we can decode back the noncommutative structure of the phase space operators in terms of the epistemic restriction and exploit it to devise classical algorithms which efficiently compute the quantum expectation values.   

Next, the transition probability in the classical algorithm of Eq. (\ref{classical computation for ensemble average general}) can be sequentially concatenated, before computing the phase space average as follows:  
\begin{eqnarray}
\braket{O}_{\{(f_T,g_T);(f_{T-1},g_{T-1});\dots;(f_1,g_1);(S ,\rho)\}}\doteq\int{\rm d}^Nq_T~{\rm d}^Np_T\dots{\rm d}^Nq_1~{\rm d}^Np_1~{\rm d}^Nq_0~{\rm d}\xi~{\rm d}^Np_0~O(p_T,q_T)\nonumber\\
\times{\mathbb P}_{\{f_T,g_T\}}(p_T,q_T|p_{T-1},q_{T-1})\times\dots\times{\mathbb P}_{\{f_1,g_1\}}(p_1,q_1|p_0,q_0){\mathbb P}_{\{S,\rho\}}(p_0,q_0|\xi)\chi(\xi).~~~
\label{classical computation for ensemble average general - quantum circuit}
\end{eqnarray} 
Assume that each transition probability takes the form of Eq. (\ref{phase space transition probability}) in which all pairs of transformation $(f_t,g_t)$ has unitary implementation $\hat{U}_t$, $t=1,\dots,T$. As long as the total transformation (in the Heisenberg picture) does not yield cubic or higher order term of momentum operator, one can proceed as before to show that the above ER classical stochastic process can be used to efficiently compute the quantum expectation value arising in the quantum circuit, i.e., we have
\begin{eqnarray}
\braket{O}_{\{(f_T,g_T);(f_{T-1},g_{T-1});\dots;(f_1,g_1);(S ,\rho)\}}=\braket{\psi|\hat{U}_1^{\dagger}\dots\hat{U}_{T-1}^{\dagger}\hat{U}_T^{\dagger}\hat{O}\hat{U}_T\hat{U}_{T-1}\dots\hat{U}_1|\psi}. 
\label{classical simulation of finite depth quantum circuit}
\end{eqnarray}
We note an interesting and important point that this result still applies even if the initial quantum observable $\hat{O}$ in Eq. (\ref{classical simulation of finite depth quantum circuit}) contains cubic or higher order terms of $\hat{p}$, and/or some of the quantum gates $\hat{U}_t$, $t=1,\dots,T$ generate cubic or higher order terms of $\hat{p}$, as long as these terms are cancelled at the end of the processes so that the final observable $\hat{O}'=\hat{U}_1^{\dagger}\dots\hat{U}_{T-1}^{\dagger}\hat{U}_T^{\dagger}\hat{O}\hat{U}_T\hat{U}_{T-1}\dots\hat{U}_1$ does not contain such term. This is comparable to the result obtained based on the $s-$ordered quasiprobability representation \cite{Rahemi-Keshari efficient simulation quantum optics}, wherein any negativity is allowed in the quasiprobability distributions associated with the initial states, and/or generated during the intermediate quantum operations, as long as the final quantum states are nonnegatively represented.  

Furthermore, bearing in mind that the above classical algorithm is developed in the form of classical stochastic processes, we can naturally generalize the algorithm to compute the quantum expectation values arising in quantum circuits which start from an arbitrary incoherent mixture (convex combination) of pure states, and evolves according to an arbitrary mixture of the (allowed class of) unitary quantum gates, as long as the associated classical mixing probabilities can be efficiently sampled. We only need to combine the sampling from the ER phase space distribution associated with the initial pure states given by Eqs. (\ref{ER phase space distribution}) and (\ref{wave function}), with the sampling from their mixing probabilities. Moreover, the transition probability must now become a mixture of the transition probabilities generated by each unitary gate. Hence, incoherent mixing does not present significant fundamental difference. 

Finally, let us discuss the convergence rate of the above classical algorithm for computing quantum expectation values using the Monte Carlo sampling from the ER phase space distribution. For this purpose, let us make more explicit the steps of the classical algorithm. For the sake of discussion, first, we ignore the unitary transformation and consider the computation of the quantum expectation value $\braket{\psi|\hat{O}|\psi}$ of a Hermitian observable $\hat{O}$ over a pure quantum state $\ket{\psi}$. Our classical algorithm starts by sampling the phase space variables $(p,q)$ from the ER phase space distribution ${\text P}_{\{S,\rho\}}(p,q|\xi)$ associated with $\psi(q)=\braket{q|\psi}=\sqrt{\rho(q)}e^{iS(q)/\hbar}$ given by Eq. (\ref{ER phase space distribution}). We then proceed to evaluate $O(q,p)$ associated with $\hat{O}$ for the above sampled value of $(p,q)$. We independently repeat this protocol for sufficiently large $K_{\rm C}$ number of times. Let us denote the value of $O$ for the $k$th sample as $\tilde{O}_k$, and compute the classical sample mean (average) 
\begin{eqnarray}
\braket{O}_{\{(S,\rho);K_{\rm C}\}}\doteq\frac{1}{K_{\rm C}}\sum_{k=1}^{K_{\rm C}}\tilde{O}_k. 
\label{classical sample mean}
\end{eqnarray}
The law of large numbers and Eq. (\ref{equivalence principle}) then ensures that in the limit of infinite number of samples, the above classical sample mean approaches the quantum expectation value as 
\begin{eqnarray}
\lim_{K_{\rm C}\rightarrow\infty}\braket{O}_{\{(S,\rho);K_{\rm C}\}}=\int {\rm d}^Nq~{\rm d}\xi~{\rm d}^Np~O(q,p){\mathbb P}_{\{S ,\rho\}}(p,q|\xi)\chi(\xi)=\braket{\psi|\hat{O}|\psi}.
\label{classical sample mean approaches true mean}
\end{eqnarray}

We note that $\tilde{O}_k$, i.e., the value of the physical quantity $O(p,q)$ computed for the $k$th sample, is in general not equal to one of the eigenvalues of the associated quantum observable $\hat{O}$. $O(p,q)$ is unbounded as can be seen from the computation of $p$ in Eq. (\ref{computation of initial momentum}), and may take on any continuum real number even when the associated quantum observable $\hat{O}$ only allows discrete spectrum of eigenvalues. Moreover, as discussed above, we can straightforwardly insert a quantum circuit generating a unitary transformation $\hat{U}_{\{f,g\}}$ as long as the resulting transformed observable $\hat{O}'=\hat{U}^{\dagger}_{\{f,g\}}\hat{O}\hat{U}_{\{f,g\}}$ is at most second order in $\hat{p}$, and the associated mapping of the classical quantity $O(p,q)\mapsto O'(p,q)$ can be obtained via a phase space mapping, $q\mapsto f(p,q)$ and $p\mapsto g(p,q)$, that is computationally trackable using classical computer.   

How does the classical sample mean $\braket{O}_{\{(S,\rho);K_{\rm C}\}}$ for a finite $K_{\rm C}$ number of samples computed in Eq. (\ref{classical sample mean}) approaches the targeted quantum expectation value $\braket{\psi|\hat{O}|\psi}$, or, how many samples $K_{\rm C}$ are required so that the classical sample mean is within a tolerated small error from the quantum expectation value, with a sufficiently high probability? This important question on convergence rate of the estimation of the quantum expectation value by classical sample mean can be assessed by applying the Chebyshev's inequality \cite{Papoulis and Pillai book}. Namely, the probability that the classical sample mean is within an error bound $\epsilon\ge 0$ from the true quantum expectation value, is given by 
\begin{eqnarray}
{\mathbb P}\Big(\big|\braket{O}_{\{(S,\rho);K_{\rm C}\}}-\braket{\psi|\hat{O}|\psi}\big|\le\epsilon\Big)\ge 1-\frac{{\rm Var}_{\{S,\rho\}}[O]}{K_{\rm C}\epsilon^2}. 
\label{Chebyshev's inequality statistical model}
\end{eqnarray}
Here, ${\rm Var}_{\{S,\rho\}}[O]$ is the variance of the classical quantity $O$ associated with the quantum observable $\hat{O}$, over the ER phase space distribution associated with the quantum wave function $\psi(q)=\sqrt{\rho(q)}e^{iS(q)/\hbar}$, defined as 
\begin{eqnarray}
{\rm Var}_{\{S,\rho\}}[O]&\doteq&\braket{(O-\braket{O}_{\{S,\rho\}})^2}_{\{S,\rho\}}. 
\label{restricted classical variance}
\end{eqnarray}
We shall refer to this quantity as the ER classical variance, and assume that it is finite. Equation (\ref{Chebyshev's inequality statistical model}) shows that, the number of samples $K_{\rm C}$ needed for the classical sample mean estimate $\braket{O}_{\{(S,\rho);K_{\rm C}\}}$ to be within an error $\epsilon$ from the true quantum expectation value $\braket{\psi|\hat{O}|\psi}$, with a probability at least $1-\delta$, $\delta\ge 0$, is proportional to the ER classical variance as: 
\begin{eqnarray}
K_{\rm C}\sim\frac{{\rm Var}_{\{S,\rho\}}[O]}{\epsilon^2\delta}. 
\label{number of required samples}
\end{eqnarray}
It is therefore instructive to study how the ER classical variance varies with the quantum states and the quantum observables, and how it is related to the epistemic restriction and the other signatures of nonclassicality. We leave this important and interesting problem for future study.  

\section{ Comparison with the classical simulation based on quasiprobability representation} 

We note that our strategy to classically compute the quantum expectation values arising in a class of CV quantum circuits employing the ER phase space representation is close in spirit to the classical simulation of quantum algorithm using the quasiprobability phase space representations reported in Refs. \cite{Veitch efficient simulation nonnegative WF 1,Veitch efficient simulation nonnegative WF 2,Mari efficient simulation nonnegative WF,Pashayan efficient estimation nonnegative quasiprob,Rahemi-Keshari efficient simulation quantum optics}. Both methods transform the quantum circuits or algorithms into classical albeit unconventional stochastic processes and apply the classical sampling method to evaluate the resulting multidimensional integral on classical computers. Below we compare the two schemes. 

First, we emphasize that the epistemic restriction in classical phase space captures directly the quantum uncertainty relation \cite{Agung-Daniel model,Agung ERPS representation}, and the ER phase space distribution defined in Eq. (\ref{ER phase space distribution}) is always non-negative. By contrast, the construction of quasiprobability phase space representation relies heavily on the abstract space of linear Hermitian operators, and the associated quasiprobability phase space distributions may yield negative value or highly irregular so that in general they cannot be seen as proper probabilities. Such negative quasiprobability, which is seen as the signature of nonclassicality, is difficult to grasp and its relation with the quantum uncertainty is not immediately clear. We note that negative quasiprobability is formally related with the fact that the transformation that maps the quantum states and operations to the quasiprobability distributions is (bi)linear in the state vector $\ket{\psi}$ \cite{Montina theorem negative quasiprobability,Spekkens negativity-contextuality,Ferrie theorem negative quasiprobability}. In contrast to this, as expressed in Eqs. (\ref{ER phase space distribution}) and (\ref{wave function}), the ER phase space distribution is nonlinear in $\ket{\psi}$. 

Further, recall importantly that the classical algorithm based on the ER phase space representation can only efficiently compute the quantum expectation values arising in a class of quantum circuits with the final quantum observables at most second order in the momentum operators (Theorem 4). Namely, it cannot be used to compute the expectation values of arbitrary Hermitian operators, including the expectation values of arbitrary POVM giving the probability of measurement outcomes. This limitation is in a sense to be expected. If our classical algorithm would apply to all forms of quantum observables, then quantum speedup might not exist. Additionally, notice that Eqs. (\ref{equivalence principle}) or (\ref{classical computation for ensemble average general}) are basically local hidden variable models for quantum expectation values, so that, according to Bell theorem for continuous variable \cite{Cavalcanti Bell inequality CV}, they cannot be applied to compute the expectation values of arbitrary Hermitian operators. On the other hand, within the quasiprobability approach, one can in principle directly use the classical sampling algorithm to efficiently simulate the probability of measurement outcomes as long as the quasiprobability distributions associated with the initial quantum states, quantum operations, and POVM are all nonnegative \cite{Veitch efficient simulation nonnegative WF 1,Veitch efficient simulation nonnegative WF 2,Mari efficient simulation nonnegative WF,Pashayan efficient estimation nonnegative quasiprob,Rahemi-Keshari efficient simulation quantum optics}. In this sense, the negativity in the quasiprobability representation was then suggested as a nonclassical ingredient responsible for the quantum speedup \cite{Veitch efficient simulation nonnegative WF 1,Mari efficient simulation nonnegative WF,Howard contextuality magic,Delfosse contextuality as resource}. With this fact in mind, let us confine our discussion to the computation of the quantum expectation values arising in a class of CV quantum circuits yielding observables at most second order in the momentum operators so that both the classical algorithm based on the ER phase space representation and those based on the quasiprobability representation apply. In this case, as mentioned in Theorems 2 and 4, unlike the classical algorithms based on the quasiprobability approach using the Wigner function reported in Refs. \cite{Veitch efficient simulation nonnegative WF 2,Mari efficient simulation nonnegative WF}, the classical algorithm based on the ER phase space representation can be applied to a large class of such quantum circuits with the initial quantum states admitting negative Wigner function and with nonGaussian quantum operations inducing nonlinear transformations. Our results thus show that, for the specific class of quantum circuits computing the quantum expectation values with the final quantum observables in the Heisenberg picture at most second order in the momentum operators, negativity of Wigner functions and nonlinearity of quantum transformation are not sufficient for quantum speedup.       

Next, employing the general quasiprobability representation developed in Ref. \cite{Ferrie-Emerson quasiprobability from frame}, Pashayan {\it et al}. in Ref. \cite{Pashayan efficient estimation nonnegative quasiprob} devised a classical sampling algorithm which mitigates the presence of negativity problem in quasiprobability approach. They showed that the classical simulation converges at a slower rate as the amount of the negativity in the quasiprobability distributions increases. Note however that, as for the general classical simulation approach based on quasiprobability representations, to run the algorithm, one needs to compute the associated quasiprobability distributions. For CV systems of large size, unless the initial quantum state and operations are factorizable into those of smaller systems, this computation of the quasiprobability distributions typically involves convoluted multidimensional integrations, which are in general computationally hard. In this sense, these quasiprobability distributions are in general difficult to sample. Moreover, to construct the classical algorithm, one needs to compute the total amount of negativity in the quasiprobability distributions which is also difficult to do for large nonfactoziable systems. The total amount of negativity is crucial in the convergence analysis of the Monte Carlo sampling technique. By contrast, as can be seen in Eq. (\ref{ER phase space distribution}), to get the associated ER phase space distribution ${\mathbb P}_{\{S,\rho\}}(p,q|\xi)$ from the initial wave function $\psi(q)=\sqrt{\rho(q)}e^{iS(q)/\hbar}$, we only need to perform spatial differentiations locally around each sample trajectory, which are in general easier to handle. The classical algorithm based on the ER phase space representation indeed requires that $\psi(q)=\braket{q|\psi}$ can be computed efficiently, $\rho(q)=|\psi(q)|^2$ can be sampled efficiently, and the mapping of Eq. (\ref{classical deterministic phase-space mapping}) is trackable on classical computers. These requirements are arguably less restrictive than the requirement of product form initial quantum states and operations in the quasiprobability approach. Furthermore, the convergence rate is determined by the ER classical variance defined in Eq. (\ref{restricted classical variance}) which can also be estimated using the Monte Carlo sampling technique.  

\section{\bf Conclusion and discussion}

Despite many continuous efforts and remarkable progresses in the last decades, the boundary and correspondence between classical and quantum computations is still not fully understood. This practically important basic problem in quantum computation is arguably deeply related to the longstanding foundational problem of quantum-classical divide. In this work, we employ a novel ER phase phase representation of quantum statistics \cite{Agung-Daniel model,Agung ERPS representation}, to devise a classical algorithm that can efficiently compute the quantum expectation values arising in a class of CV quantum circuits which yield, after the Heisenberg time evolution, quantum observables at most second order in momentum. The classical algorithm is obtained by transforming the quantum circuits in the complex Hilbert space into classical stochastic processes in the classical phase space, but with the initial positions and momentum being sampled from a specific ER phase space distribution associated with the initial quantum state. We then use the Monte Carlo sampling technique to evaluate the resulting multidimensional integral. It remains an open challenging important problem to extend the method to discrete variable quantum computations.  

The classical algorithm shows that the epistemic restriction, including the nonseparability of $\xi$, is necessary for quantum algorithms computing the quantum expectation value. Namely, we can define quantum computations as those that benefits from a specific epistemic restriction in the classical phase space, encoded as canonical commutation relation in the abstract Hilbert space. In the context of computation, the epistemic restriction may thus be (paradoxically) liberating. Moreover, for a certain class of quantum circuits computing the expectation values, we can decode the canonical commutation relation back in the form of epistemic restriction in the classical phase space, and exploit it to devise an efficient classical algorithm to compute the quantum expectation values. The result suggests that the epistemic restriction may offer an intuitive conceptual tool to further study the boundary between quantum and classical computations. To this end, it is interesting to explore the general connection between the epistemic restriction and other signatures of nonclassicality suggested to bound the convergence rate of classical sampling algorithms, e.g., negativity in quasiprobability approach \cite{Pashayan efficient estimation nonnegative quasiprob}, and interference in an approach based on sampling over Feynman-like paths \cite{Stahlke efficient estimation interference}.  

On the other hand, the epistemic restriction is not sufficient for all possible (universal) quantum computational algorithms. The classical algorithm of Eq. (\ref{classical computation for ensemble average general}) does not simulate the dynamics of the quantum states (except for the Gaussian sector as mentioned in Corollary 3), and the random outcome in quantum measurement. As discussed in Ref. \cite{Agung-Daniel model}, to recover the dynamics of the quantum states within the ER phase space representation, we need to follow the dynamics of the whole ensemble of trajectories which is required to respect the conservation of average energy. Moreover, to get a definite measurement outcome, we also need to keep track of each single random trajectory which is allowed to violate the conservation of energy. This observation suggests that the key behind the quantum speedup might be the ability of Nature to manage conserving the ensemble average energy in the presence of a global random variable, while allowing each trajectory in the ensemble to violate randomly the conservation of energy. Intuitively, to manage to do these dual tasks naively using only classical resource is computationally hard. By contrast, in classical mechanics, the average ensemble energy is automatically conserved since each single trajectory is deterministic conserving the energy. As an analogy, while each trajectory in the Bohmian mechanics \cite{Bohmian mechanics} violates the conservation of (local) energy (defined suitably to include the so-called quantum potential), the whole ensemble of Bohmian trajectories manage to mysteriously anticipate each other so that the ensemble average energy is conserved. In Bohmian mechanics, this is due to the presence of a physical wave function evolving under the Schr\"odinger equation, which co-orchestrates the whole ensemble of trajectories. We hope to further clarify this conjecture in the future.

\bigskip
{\bf Acknowledgements}\\
A. B. acknowledges the supports from the John Templeton Foundation (project ID 43297). The opinions expressed in this publication do not necessarily reflect the views of the John Templeton foundation. The work is also partially funded by the Ministry of Education and Culture, and the Ministry of Research and Technology of Republic of Indonesia, under the grant scheme ``Penelitian Dasar Unggulan Perguruan Tinggi (PDUPT),'' and the WCU Program managed by Institut Teknologi Bandung. A.B. would like to thank Daniel Rohrlich for encouraging advice and continuous support in the present research, and Joseph Emerson for the stimulating discussion during the initial phase of the present work. The Authors also thank the Referees for the constructive comments and recommendations.\\


\begin{thebibliography}{10}  

\bibitem{Feynman quantum computer} Feynman, R. P. Simulating physics with computers. {\it International Journal of Theoretical Physics} 21, 467-487 (1982). 
\bibitem{Lloyd quantum simulator} Lloyd, S. Universal quantum simulators. {\it Science} 273, 1073-1078 (1996). 
\bibitem{Shor factorization algorithm} Shor, P. Polynomial-time algorithms for prime factorization and discrete logarithms on a quantum computer. {\it SIAM Journal on Computing} 26, 1484-1509 (1997).

\bibitem{Gottesman-Knill theorem} Gottesman, D. The Heisenberg representation of quantum computers. ArXiv:quant-ph/9807006 (1998).
\bibitem{Nielsen-Chuang book} Nielsen, M. A. \& Chuang, I. L. Quantum computation and quantum information. (Cambridge University Press, Cambridge, 2000).
\bibitem{Aaronson GK theorem improved} Aaronson, S. \& Gottesman, D. Improved Gottesman-Knill theorem. {\it Phys. Rev. A} 70, 052328 (2004).

\bibitem{Bartlett continuous variable GK theorem} Bartlett, S. D., Sander, B. C., Braunstein, S. L. \& Nemoto, K. Efficient classical simulation of continuous variable quantum information processes. {\it Phys. Rev. Lett.} 88, 097904 (2002). 
\bibitem{Bartlett efficient simulation Gaussian quantum optics} Bartlett, S. D. \& Sanders, B. C. Efficient classical simulation of optical quantum information circuits. {\it Phys. Rev. Lett.} 89, 207903 (2002). 

\bibitem{Veitch efficient simulation nonnegative WF 1} Veitch, V., Ferrie, C., Gross, D. \& Emerson, J. Negative quasi-probability as a resource for quantum computation. {\it New J. Phys.} 14, 113011 (2012).
\bibitem{Veitch efficient simulation nonnegative WF 2} Veitch, V., Wiebe, N., Ferrie C. \& J. Emerson. Efficient simulation scheme for a class of quantum optics experiments with non-negative Wigner phase space representation. {\it New J. Phys.} 15, 013037 (2013).
\bibitem{Mari efficient simulation nonnegative WF} Mari, A. \& Eisert, J. Positive Wigner function renders classical simulation of quantum computation efficient. {\it Phys. Rev. Lett.} 109, 230503 (2012).
\bibitem{Pashayan efficient estimation nonnegative quasiprob} Pashayan, H., Wallman, J. J. \& Bartlett, S. D. Estimating outcome probabilities of quantum circuits using quasiprobabilities. {\it Phys. Rev. Lett.} 115, 070501 (2015).
\bibitem{Rahemi-Keshari efficient simulation quantum optics} Rahemi-Keshari, S., Ralph, T. C. \& Caves, C. M. Sufficient conditions for efficient classical simulation of quantum optics. {\it Phys. Rev. X} 6, 021039 (2016). 

\bibitem{Valiant efficient simulation Match gates} Valiant, L. G. Quantum computer that can be simulated classically within polynomial time. {\it STOC '01 Proceedings of the thirty-third annual ACM symposium on theory of computing}, pp. 114-123 (2001).
\bibitem{Knill efficient simulation fermionic linear optics} Knill, E. Fermionic linear optics and match gates. ArXiv:quant-ph/0108033 (2001). 
\bibitem{Terhal efficient simulation fermionic linear optics} Terhal, B. \& DiVincenco, D. P. Clasical simulation of noninteracting-fermion quantum circuits. {\it Phys. Rev. A} 65, 032325 (2002). 
\bibitem{Jozsa efficient simulation match gates} Jozsa, R. \& Miyake A. Match gates and classical simulation of quantum circuits. {\it Proc. R. Soc. A} 464, 3089-3106 (2008).

\bibitem{Vidal efficient simulation small entanglement} Vidal, G. Efficient classical simulation of slightly entangled quantum computations. {\it Phys. Rev. Lett.} 91, 147902 (2003). 
\bibitem{Jozsa efficient simulation small entanglement} Jozsa R. \& Linden N. On the role of entanglement in quantum-computational speedup. {\it Proc. R. Soc. Lond.} A 459, 2011-2032 (2003). 
\bibitem{Shi efficient simulation small entanglement} Shi, Y. -Y., Duan, L. M. \& Vidal, G. Classical simulation of quantum many-body systems with a tree tensor network. {\it Phys. Rev. A} 74, 022320 (2006). 

\bibitem{Nest MBQC efficient simulation small entanglement} Van den Nest, M., D\"ur, W., Vidal,  G. \& Briegel, H. J. Classical simulation versus universality in measurement based quantum computation. {\it Phys. Rev. Lett.} 97, 150504 (2006).
\bibitem{Gross MBQC efficient simulation too much entanglement} Gross, D., Flammia, S. T. \& Eisert, J. Most quantum states are too entangled to be useful as computational resources. {\it Phys. Rev. Lett.} 102, 190501 (2009).

\bibitem{Yoran efficient simulation limited width circuits} Yoran, N. \& Short, A. J. Classical simulation of limited width cluster-state quantum computation. {\it Phys. Rev. Lett.} 96, 170503 (2006).
\bibitem{Markov efficient simulation limited width circuits} Markov, I. L. \& Shi, Y. Simulating quantum computation by contracting tensor networks. {\it SIAM Journal on Computing} 38(3), 963-981 (2008). 
\bibitem{Jozsa efficient simulation limited width circuits} Jozsa, R. On the simulation of quantum circuits. ArXiv:quant-ph/0603163 (2006). 
\bibitem{Garcia-Alvarez efficient simulation of QC with large negative Wigner function} Garc\'ia-\'Alvarez, L., Calcluth, C., Ferraro, A. \& Ferrini, G. Efficient simulatability of continuous-variable circuits with large Wigner negativity. ArXiv:quant-ph/2005.12026v1 (2020).

\bibitem{Veitch resource theory quantum computation} Veitch, V., Mousavian, S. A. H., Gottesman, D. \& Emerson, J. The resource theory of stabilizer computation. {\it New J. Phys.} 16, 013009 (2014).
\bibitem{Howard contextuality magic} Howard, M., Wallman, J., Veitch, V. \& Emerson, J. Contextuality supplies the `magic' for quantum computation. {\it Nature} 510, 351-355 (2014). 
\bibitem{Delfosse contextuality as resource} Delfosse, N., Guerin, P. A., Bian, J. \& Raussendorf, R. Wigner function negativity and contextuality in quantum computation on rebits. {\it Phys. Rev. X} 5, 021003 (2015). 
\bibitem{Vedral quantum computational resource} Vedral, V. The elusive source of quantum speedup. {\it Found. Physics} 40, 1141-1154 (2010).

\bibitem{Aaronson BosonSampling} Aaronson, S. \& Arkhipov, A. The computational complexity of linear optics. {\it Theory of Computing} 9(4), 143-252 (2013).
\bibitem{Bremmer quantum supremacy IQP} Bremner, M. J., Jozsa, R. \& Shepherd, D. J. Classical simulation of commuting quantum computations implies collapse of the polynomial hierarchy. {\it Proc. R. Soc. A} 467, 459-472 (2010).
\bibitem{Harrow quantum supremacy review} Harrow, A. W. \& Montanaro, A. Quantum computational supremacy. {\it Nature} 549, 203-209 (2017).
\bibitem{Lund quantum supremacy review} Lund, A. P., Bremner, M. J. \& Ralph, T. C. Quantum sampling problems, BosonSampling and quantum supremacy. {\it npj Quantum Information} 3:15 (2017). 
\bibitem{Boixo characterizing quantum supremacy} Boixo, S., Isakov, S. V., Smelyanskiy, V. N., Babbush, R., Ding, N., Jiang, Z., Bremner,  M. J., Martinis, J. M., John, M. \& Neven, H. Characterizing quantum supremacy in near-term devices. {\it Nature Physics} 14 (6), 595-600 (2018). 

\bibitem{Hardy epistricted model} Hardy, L. Disentangling nonlocality and teleportation. ArXiv:quant-ph/9906123 (1999).
\bibitem{Emerson PhD thesis} Emerson, J. V., Quantum chaos and quantum-classical correspondence, PhD thesis,  Simon Fraser University (2001).
\bibitem{Spekkens toy model with epistemic restriction} Spekkens, R. W. Evidence for the epistemic view of quantum states: a toy theory. {\it Phys. Rev. A} 75, 032110 (2007). 
\bibitem{Bartlett reconstruction of Gaussian QM with epistemic restriction} Bartlett, S. D., Rudolph, T. \& Spekkens, R. W. Reconstruction of Gaussian quantum mechanics from Liouville mechanics with an epistemic restriction. {\it Phys. Rev. A} 86, 012103 (2012).

\bibitem{Agung-Daniel model} Budiyono, A. \& Rohrlich, D. Quantum mechanics from classical statistical mechanics with an epistemic restriction and an ontic extension. {\it Nature Communications} 8, 1306  (2017).
\bibitem{Agung ERPS representation} Budiyono, A. Epistemically restricted phase-space representation, weak momentum value, and reconstruction of the quantum wave function. {\it Phys. Rev. A} 100, 032125 (2019). 

\bibitem{Agung epistemic interpretation} Budiyono, A. Quantum mechanics as a calculus for estimation under epistemic restriction. {\it Phys. Rev. A} 100, 062102 (2019).  
\bibitem{Agung estimation independence} Budiyono, A. Estimation independence as a physical principle for quantum uncertainty. {\it Phys. Rev. A} 101, 022102 (2020). 
\bibitem{Agung nonlinear Schroedinger equation and estimation independence} Budiyono, A. \& Dipojono, H. K. Nonlinear Schr\"odinger equations and generalized Heisenberg uncertainty principle from estimation schemes violating the principle of estimation independence. {\it Phys. Rev. A }102, 012205 (2020). 

\bibitem{Lloyd continuous variable UQC} Lloyd, S. \& Braunstein, S. L. Quantum computation over continuous variables. {\it Phys. Rev. Lett.} 82, 1784-1787 (1999).  
\bibitem{Gottesman continuous variable UQC} Gottesman, D., Kitaev,  A. \& Preskill, J. Encoding a qubit in a harmonic oscillator. {\it Phys. Rev. A} 64, 012310 (2001).   

\bibitem{Cerezo review VQA} Cerezo, M., Arrasmith, A., Babbush, R., Benjamin, S. C., Endo, S., Fujii, K., McClean, J. R., Mitarai, K., Yuan, X., Cincio, L. \& Coles, P. J. Variational Quantum Algorithms. arXiv:2012.09265v1 (2020). 

\bibitem{Hillery quasiprobability review} Hillery, M., O'Connel,  R. F., Scully, M. O. \& Wigner, E. P. Distribution functions in physics: fundamentals. {\it Phys. Rep.} 106, 121-167 (1984).
\bibitem{Lee quasiprobability review} Lee, H-W. Theory and application of the quantum phase-space distribution functions. {\it Physics Reports} 259, 147-211 (1995).
\bibitem{Ferrie quasiprobability review} Ferrie, C. Quasi-probability representations of quantum theory with applications to quantum information science. {\it Report on Progress in Physics} 74, 116001 (2011).

\bibitem{Rund book: Hamilton-Jacobi formalism} Rund, H. The Hamilton-Jacobi theory in the calculus of variations: its role in mathematics and physics. (Van Nostrand, London, 1966).

\bibitem{Aharonov weak value} Aharonov, Y., Albert, D. Z. \& Vaidman, L. How the result of a measurement of a component of the spin of a spin-$1/2$ particle can turn out to be $100$. {\it Phys. Rev. Lett.} 60 (14), 1351-1354 (1988).  
\bibitem{Lundeen complex weak value} Lundeen, J. S. \& Resch, K. J. Practical measurement of joint weak values and their connection to the annihilation operator. {\it Phys. Lett. A} 334, 337-344 (2005).
\bibitem{Jozsa complex weak value} Jozsa, R. Complex weak values in quantum measurement. {\it Phys. Rev. A}  76, 044103 (2007). 

\bibitem{Nest probabilistic method} Van den Nest, M. Simulating quantum computers with probabilistic methods. {\it Quantum Information} \& {\it Computation} 11, 784-812 (2011).

\bibitem{Arvind symplectic geometry} Arvind, Dutta, B., Mukunda, N. \& Simon, R. The real symplectic groups in quantum mechanics and optics. {\it Pramana} 45, 471-497 (1995).
\bibitem{Adesso CV quantum information} Adesso, G., Ragy, S. \& Lee, A. R. Continuous variable quantum information: Gaussian states and beyond. {\it Open Systems} \& {\it Information Dynamics} 21, 1440001 (2014).

\bibitem{Weedbrook Gaussian quantum information} Weedbrook, C., Pirandola, S., Garcia-Patron, R., Cerf, N. J., Ralph, T. C., Shapiro, J. H. \& Lloyd, S. Gaussian quantum information. {\it Reviews of Modern Physics} 84, 621-669 (2012). 

\bibitem{Papoulis and Pillai book} Papoulis, A. \& Pillai, S. U. Probability, random variable and stochastic processes. (McGraw-Hill, Singapore, 2002).

\bibitem{Montina theorem negative quasiprobability} Montina, A. A condition for any realistic theory of quantum systems. {\it Phys. Rev. Lett.} 97, 180401 (2006).
\bibitem{Spekkens negativity-contextuality} Spekkens, R. W. Negativity and contextuality are equivalent notions of nonclassicality. {\it Physical Review Letters} 101 (2), 020401 (2008).
\bibitem{Ferrie theorem negative quasiprobability} Ferrie, C. \& Emerson, J. Frame representations of quantum mechanics and the necessity of negativity in quasi-probability representations. {\it Journal of Physics A: Mathematical and Theoretical} 41 (35), 352001 (2008).

\bibitem{Cavalcanti Bell inequality CV} Cavalcanti, E. G., Foster, C. J., Reid, M.D. \& Drummond, P. D. Bell Inequalities for Continuous-Variable Correlations. {\it Phys. Rev. Lett.} 99, 210405 (2007). 

\bibitem{Ferrie-Emerson quasiprobability from frame} Ferrie, C. \& Emerson, J. Framed Hilbert space: hanging the quasi-probability pictures of quantum theory. {\it New Journal of Physics} 11 (6), 063040 (2009).

\bibitem{Stahlke efficient estimation interference} Stahlke, D. Quantum interference as a resource for quantum speedup. {\it Phys. Rev. A} 90, 022302 (2014). 

\bibitem{Bohmian mechanics} Bohm, D. A suggested interpretation of the quantum theory in terms of ``hidden'' variables I. {\it Phys. Rev} 85, 166-179 (1952).

\end{thebibliography}
\end{document}